# Improvement on Extrapolation of Species Abundance Distribution Across Scales from Moments Across Scales*

Saeid Alirezazadeh**,a, Khadijeh Alibabaei[b]

[a] CIBIO, InBIO, CEABN, Lisboa, Portugal, [b]Departamento de Matematica, Faculdade de Ciencias, Universidade do Porto, Porto, Portugal

**Abstract**

Raw moments are used as a way to estimate species abundance distribution. The almost linear pattern of logarithmic transformation of raw moments across scales allows us to extrapolate species abundance distribution for larger areas. However, the results may contain errors. Some of these errors are due to computational complexity, pattern matching, binning methods, etc. We present some strategies to reduce some of the errors. The main result is the introduction of new techniques to evaluate more accurate species abundance distributions across scales through moments across scales.

## 1. Introduction

Species abundance distribution (SAD) describes the relationship between the number of observed species in a sample and their relative abundance. It represents the full distribution of rarity and commonness in sample data. The SAD is one of the most widely studied patterns in ecology. Therefore, several different models are used to characterize the shape of the distribution and identify the potential mechanisms of the pattern. Fisher et al. (1943) proposed that logseries is the theoretical distribution for the relative abundance of species. Preston (1948), by looking at SADs at different scales, suggested instead the lognormal distribution as the limiting distribution. Bulmer (1974 and 1979) showed that the Poisson lognormal is a better alternative to the lognormal. More recently, Engen and Lande (1996) also considered the compounded Poisson Gamma distribution. May (1979) stated that the logseries can be regarded as the distribution characteristic of relatively simple communities whose dynamics are dominated by a single factor. On the other hand, when the environment fluctuates randomly or multiple factors become significant, the central limit theorem yields the lognormal distribution.

* This work is done within 2017-2019. On the date of publishing on ArXiv, **Saeid Alirezazadeh** is with C4 – Cloud Computing Competence Centre (C4-UBI), Universidade da Beira Interior, Rua Marquês d'Ávila e Bolama, 6201-001, Covilhã, Portugal, Portugal, acknowledge the support given by operation Centro-01-0145-FEDER-000019 - C4 – Centro de Competências em Cloud Computing, cofinanced by the European Regional Development Fund (ERDF) through the Programa Operacional Regional do Centro (Centro 2020), in the scope of the Sistema de Apoio à Investigação Cientifíca e Tecnológica - Programas Integrados de IC&DT. **Khadijeh Alibabaei** is with C-MAST Center for Mechanical and Aerospace Science and Technologies, University of Beira Interior, Deparment of Electromechanical Engineering 6201-001, Covilhã, Portugal acknowledge the support given by the project Centro-01-0145-FEDER000017-EMaDeS-Energy, Materials, and Sustainable Development, co-funded by the Portugal 2020 Program (PT 2020), within the Regional Operational Program of the Center (CENTRO 2020) and the EU through the European Regional Development Fund (ERDF). Fundação para a Ciência e a Tecnologia (FCT—MCTES) also provided financial support via project UIDB/00151/2020 (C-MAST).
** Corresponding author E-mail address: saeid.alirezazadeh@gmail.com

Although most work on SADs has used a fixed scale, there are exceptions. Hubbell (2001) developed the neutral theory, which attempts to predict SADs across scales in space and time. This theory was called neutral because it assumed that all individuals in a community are equivalent in terms of reproductive rates, mortality, dispersal, and speciation. Two of Hubbell's speciation modes, the point mutation mode and the fission mode, resulted in SADs that varied according to scale. Although the fission mode led to the same type of distribution at different scales, the point mutation mode predicted the zero-sum multinomial distribution (a distribution introduced by Hubbell (2001)) at small scales (the local community) and predicted the logseries distribution at large scales (the metacommunity). Therefore, neutral theory acknowledges that the SAD changes across scales.

Typically, empirical SADs come from samples that represent only a small fraction of the community, namely the quantities that are practical (or economical) to obtain. If we could find a pattern for how SADs scale with the area (or other parameters), this might allow us to predict distributions for larger areas, at least within some reasonable scales that are likely to be dictated by characteristics of the landscape in which the community exists, such as habitat homogeneity.

Moment functions are used in image analysis and related applications, such as pattern recognition, object identification, template matching, and pose estimation (see Liao and Pawlak (1996), Teague (1980), and Mukundan, Ong, and Lee (2001)). However, their application to the study of SADs has only recently been attempted (see Borda-de-Água et al. (2012 and 2017) and Alirezazadeh et al. (2018)). Here, we define moments, $M_n$ (often called raw moments), of a sample as

$$M_n = \frac{1}{S}\sum_{i=1}^{S} x_i^n, \qquad (1)$$

where $n$ defines the order of the moment and $x_i$ is the $log_2$ of the number of individuals of the species $i$ in the sample, which consists of $S$ species. Moments are important because knowing them is enough to reconstruct the probability density function; for example, with $n = 1$ we obtain the mean, with $n = 2$ we get the value related to the variance, and so on, Feller (2008).

Note that all the existing samples are contained in a larger area whose species abundance distribution may not be known. For a given sample, if we could find any pattern for species abundance distributions across scales, we can estimate species abundance distribution for a larger scale and thus for the entire community in which the sample is contained.

After finding moments of different orders, we can quickly obtain Tchebychev moments directly from the moments. Using Tchebychev polynomials, which are fixed polynomials (first provided by Tchebychev (1854)), we have a polynomial estimate for the species abundance distribution. This procedure for a given set of moments has different levels of complexity which, when improved, gives a better polynomial estimate of the real distribution. The first is the computational complexity of the Tchebychev polynomials and moments, where we use the recurrence relation between polynomials of each degree and then introduce the matrix multiplier that yields the Tchebychev moments from the raw moments. The next complexity is the raw moments themselves, because the method of Tchebychev polynomials is very sensitive to the values of the raw moments. Slight changes in their values make large differences. We will use the distribution moments instead. But the distribution moments do not have a linear behavior like the raw moments by areas in log-log scale. This latter linear behavior allows extrapolation of moments for larger scales. We will provide a solution on how to extrapolate distribution moments using their relation to raw moments. Note that algorithms combining symbolic and numerical computational techniques have gained importance and interest in recent years. The need to work reliably with imprecise and noisy data, as well as speed and accuracy in algebraic and hybrid symbolic-numeric problems, has spurred the rapid development of the field of symbolic-numeric computation. new and exciting problems in industrial, mathematical and computational areas are now being explored and solved. We use a symbolic-numerical technique to reduce the computational complexity.

Our examples are the tropical forest sample Barro Colorado Island (BCI) (Hubbell et al. (2005), Hubbell et al. (1999), Condit (1998), Condit et al. (2017)) and all individuals of all species with at least 10 cm and sometimes (1 cm) (dbh) (Diameter at Breast Height) has been considered.

## 2. Preliminaries

There are several equivalent ways to formulate raw moments; the following is the most convenient one used in the examples: Let a sample data $A$ consists of individuals of different species. Let $x_i$ be the logarithm of base 2 of the total number of individuals of species $i$, and let $k$ be the total number of species in the sample. Then the (raw) moment of order $t$, denoted by $M_t(A)$, can be defined as follows:

$$M_t(A) = \frac{1}{k}\sum_{i=1}^{k} x_i^t.$$

For more details and application of moments, see Borda-de-Água, Hubbell, and McAllister (2002); Borda-de-Água et al. (2012); Borda-de-Água et al. (2017); Mukundan (2004). This allows us to examine moments across scales. In Figure 1, we plot the log-transformed moments up to twentieth order by the log of the area size from area size 2 (ha) to 50 (ha) for BCI data, one considering all individuals with 1cm dbh and the other considering all individuals with 10cm dbh. As shown in Figure 1, the near linear patterns for moments of different orders and area size can be visualized in a log-log scale.

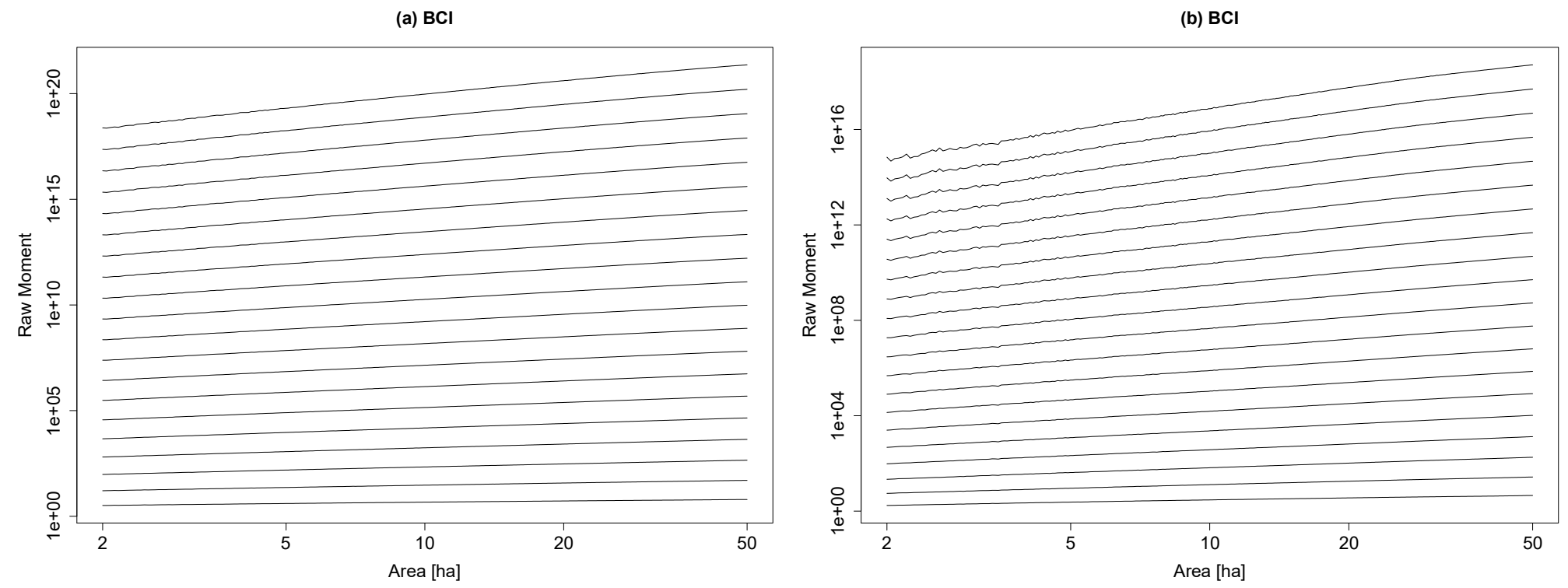

**Figure 1.** Raw moments of order 1 up to 20 as a function of the area from 2 (ha) to 50 (ha) in log-log scale for BCI by considering individuals with (a) 1cm dbh (b) 10cm dbh.

Note that raw moments are estimates of distribution moments that can be obtained from related species abundance distribution. Before arguing their differences, we need to explain the binning that causes the differences. The bins are intervals of the general form $[b, b+1)$, where $b$ takes its values from the sequence -0.5, 0.5, 1.5, 2.5, 3.5, … i.e., the bins are centered on the natural numbers 0, 1, 2, 3, …. The species $j$ is associated with the bin centered by $i$ if and only if $i$ is the smallest integer close to $\log_2(x_j)$ for $x_j$ is the number of individuals of the species $j$. Let $(x)$ denotes the smallest integer close to $x$. Therefore, the raw moment of order $n$ can be obtained from the formula $\frac{1}{S}\sum_{i=1}^{S} x_i^n$, while the distribution moment of order $n$ can be obtained from

$$M'_n = \frac{1}{S}\sum_{i=1}^{S}(x_i)^n,$$

where $x_i$ is the $\log_2$ of the number of individuals of the species $i$ and $S$ is the total number of species. Suppose we use the distribution moments to evaluate the Tchebychev moments and use them together with the Tchebychev polynomials. In this case, we can reproduce the exact SAD, while using the raw moments will reproduce an estimate for SAD.

In Figure 2, we plot the log-transformed of distribution moments up to the twentieth order by the logarithm of the area size from area size 2 (ha) to 50 (ha) again for BCI data, once considered with all individuals with 1cm dbh and the other time with all the individuals with 10 cm dbh.

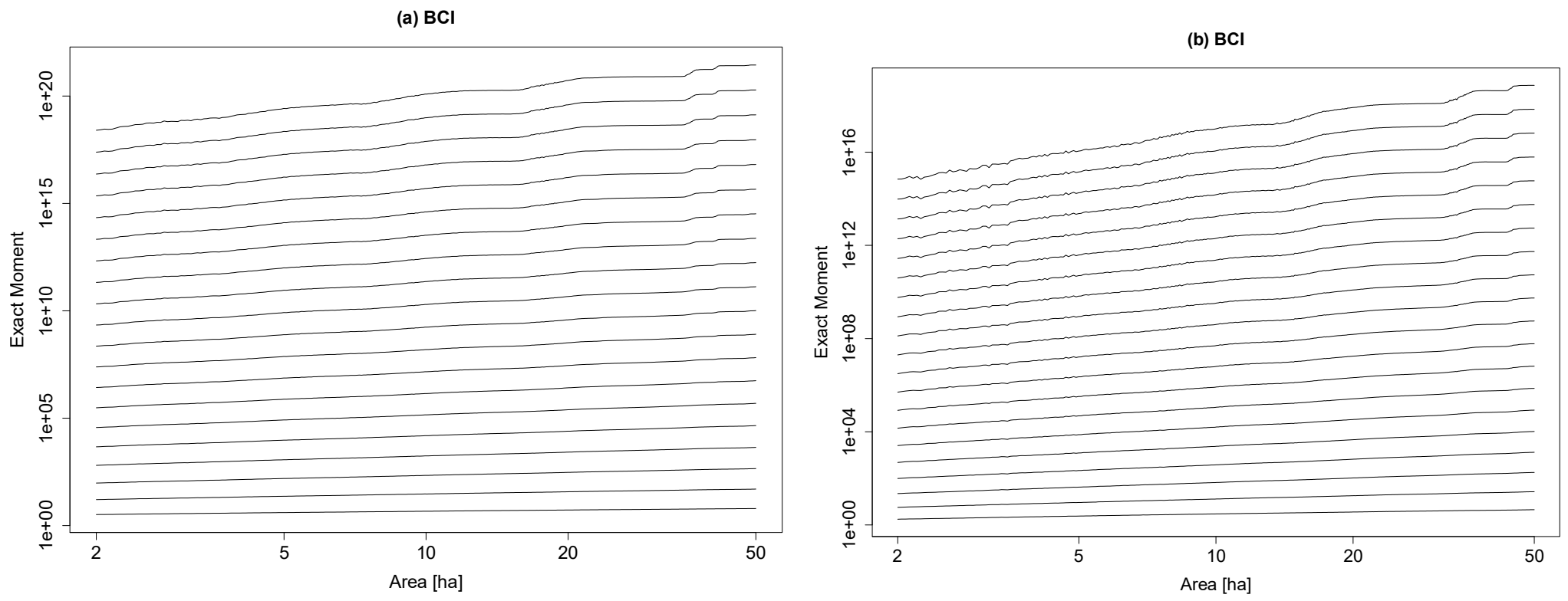

**Figure 2.** Distribution moments of order 1 up to 20 as a function of the area from 2 (ha) up to 50 (ha) in log-log scale for BCI by considering individuals with (a) 1cm dbh (b) 10cm dbh.

As we can see, the pattern of distribution moments by area in the log-log scale is not linear. We used moments up to order $20^{th}$ in order to have a better visual of their non-linear patterns across scales.

## 3. Patterns

If we consider the relation between $\log(M'_n(x))$ and $\log(M_n(x))$, it turns out that it is a linear function through 0, see Figure 3. However, slopes are changing when we change the area size. Consider the function

$$C_n(x) = slope\big(\log(M'_n(x)) \sim \log(M_n(x))\big),$$

where $x$ corresponds to the area size. The function $C_n(x)$ is defined as the slope of the linear regression between the log of the raw and log of the distribution moments at the area of size $x$. Hence, finding the functions $C_n(x)$ and $\log(M_n(x))$ allow us to find the functionality of $M'_n(x)$. This allows us to have a better estimation of the SAD by extrapolating the area.

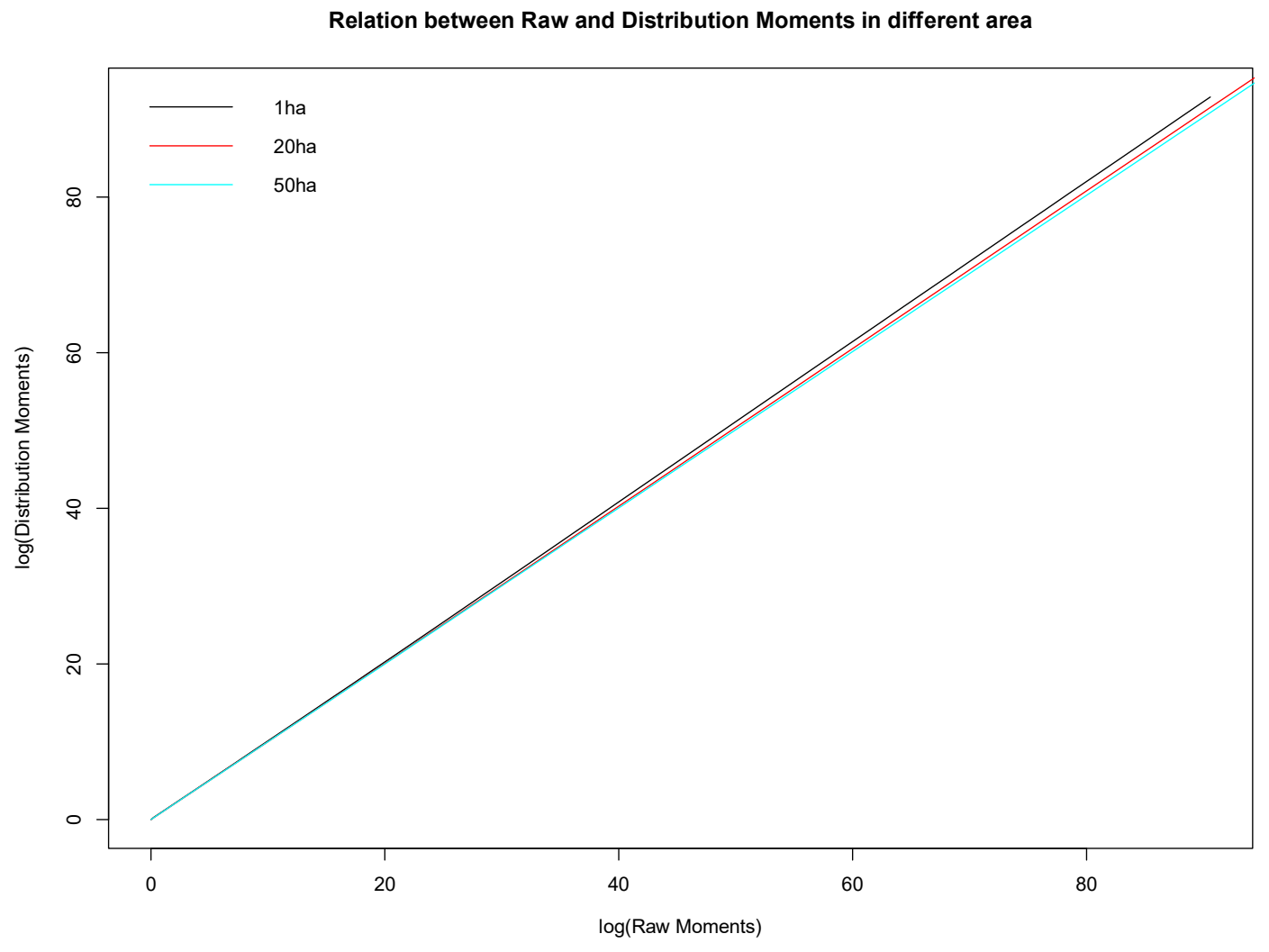

**Figure 3.** Relation between Raw and Distribution moments of different orders in 1, 20, and 50 ha area.

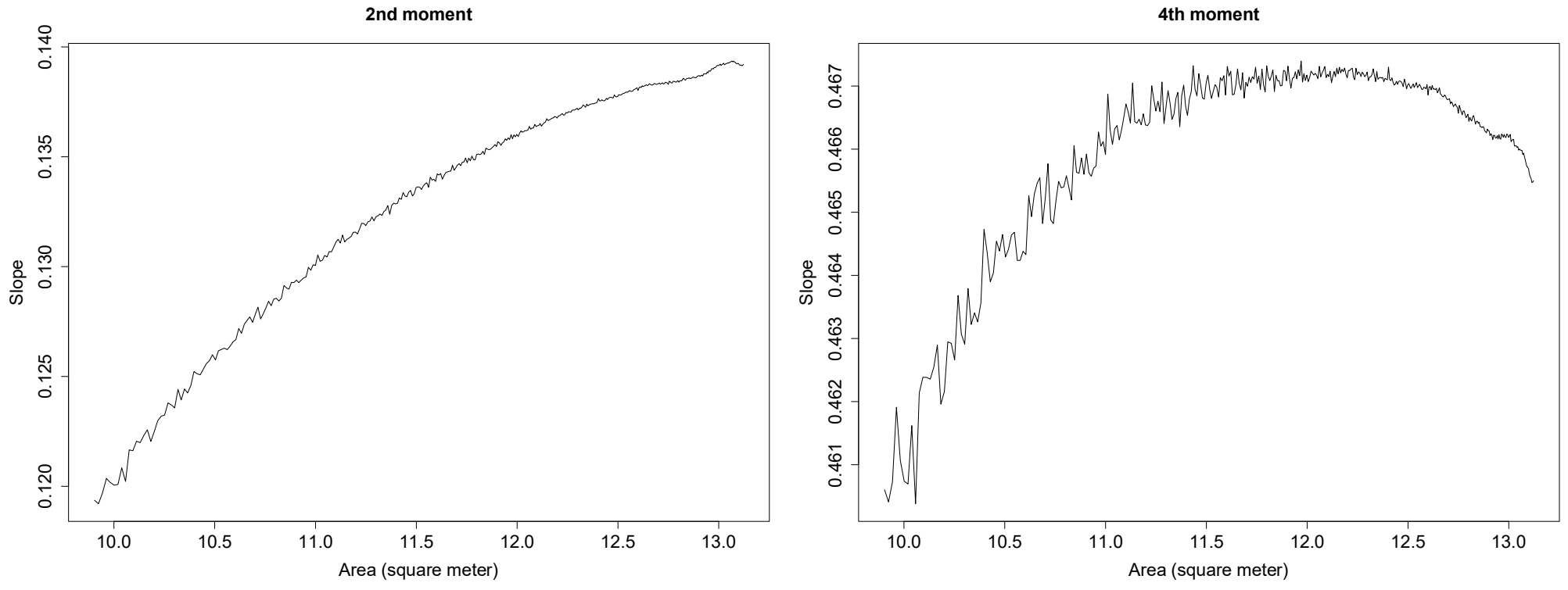

**Figure 4.** Slopes of the log of 2$^{nd}$ and 4$^{th}$ raw moments by the log of respective distribution moments across a log of area.

Figure 4 shows how the slopes' second and fourth moments change by the log of area.

### 3.1. Raw Moments

Since the almost linear patterns between the log of raw moments and log of areas happen, see Figure 1, we can extrapolate raw moments for a larger area. But as we can see in Figure 5, fitting the pattern of the log of raw moments by the log of areas as linear functions is poor. But by looking at residuals of linear fitting of raw moments, it appears that a sinusoidal function is a good choice. Hence, the patterns for the log of raw moments across scales can be obtained from the following general formula:

$$\log(M_n(x)) = a_0 + a_1 x + b \sin(cx' + \varphi),$$

where $x$ takes its values from the log of areas and $x' = \frac{x - \min Area}{\max Area - \min} \pi$ (associated values of the log of areas in the interval $[0, \pi]$). By considering the shape of the residuals, we can consider $c = 1$ and so the formula can be translated to the following non-linear function:

$$\log(M_n(x)) = a_0 + a_1 x + a_2 \sin(x') + a_3 \cos(x').$$

Hence, for the extrapolation of moments in Figure 1, we will not use linear regression. Instead, we use regression by sums of functions of a linear function of $log$ of area size, $\sin(x)$ and $\cos(x)$ where $x$ is the transformation of the log of area sizes to its respective gradient values. Then after by finding the functionality of $C_n(x)$, we can do the extrapolation of distribution moments for larger areas. Figure 5 and Figure 6 show how fitting the log of raw moments of order 10 by the log of the area with linear regression and with the preceding fitting functions are.

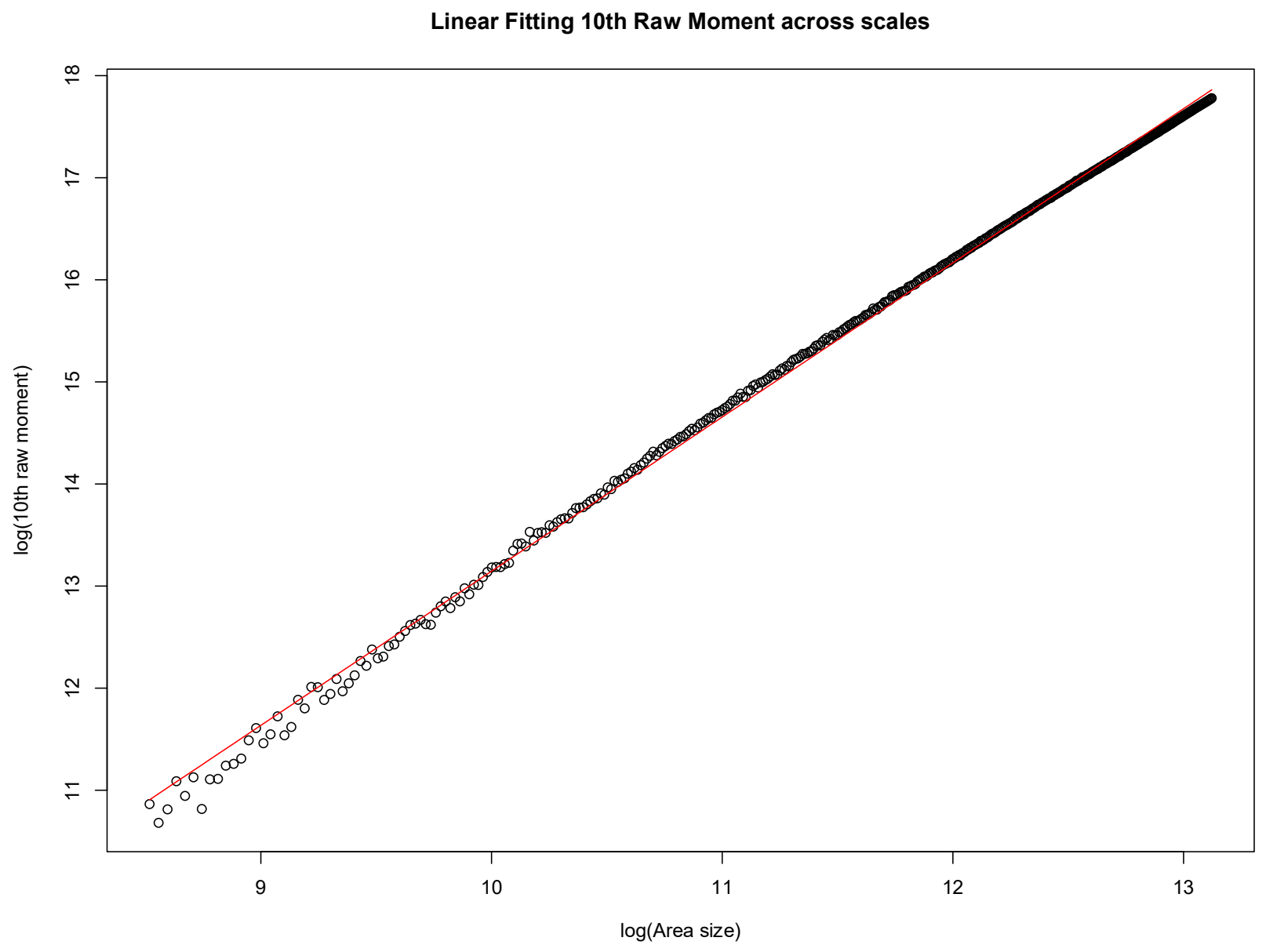

**Figure 5.** Fitting of the log of 10th raw moments by the log of areas for BCI with linear function and area size larger than 2 (ha).

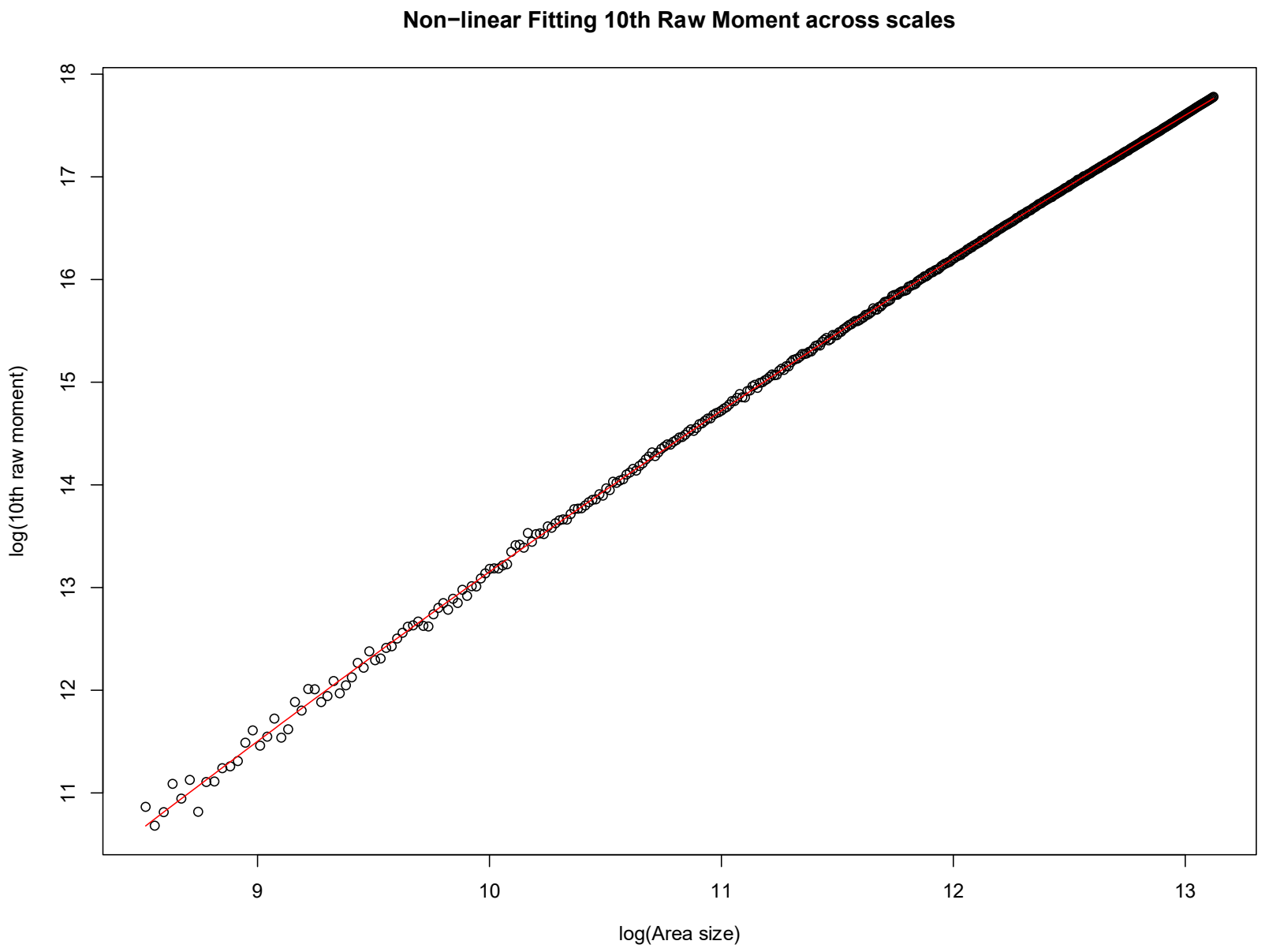

**Figure 6.** Fitting of the log of 10th raw moments by the log of areas for BCI with non-linear functions and area size larger than 2 (ha).

## 3.2. Pattern of $C_n(x)$

To find the pattern of $C_n(x)$ we need first to find and remove the possible trend in it. Note that in our examples, we considered $x$ as the square root of the area instead of area size itself, because the change of the value of $x$ must be linear. So we do not need to have additional rounding of rational or irrational numbers. For removing a possible trend, we will remove its linear regression pattern, that is:

$$C'_n(x) = C_n(x) - ax - b,$$

where $x$ corresponds to the square root of the area. This allows us to centralize the values of $C_n(x)$. From these values, we will take the $arctan$ which give us the gradient of degrees between $[-\frac{\pi}{4}, \frac{\pi}{4}]$ (values are very close to 0). Note that $\arctan(y) \approx y$ in this interval and since the change of the degrees are very small. Then recursively, we can consider the degree corresponds to extrapolated as the degree corresponds to the degree obtained from the largest area $x$.

Note that the shape of $C'_n(x)$ is related to the largest occupied bin and the number of the most abundant species, where both increase with the area. If we denote the largest occupied bin by (LB) and the number of most abundance species by (NMAS) in the area $x$, then there is a clear pattern for the shape of $C'_n(x)$ by increasing the area:

- LB and NMAS remain the same, then $C'_n(x)$ decreases;

- LB remains the same, but NMAS increases, then $C'_n(x)$ increases;

- LB increases, then $C'_n(x)$ increases.

Since there are no proper rules for values of LB and NMAS across scales, we will consider our simplified approach, which is: by increasing slightly the area $x$ from $x_1$to $x_2$ more likely, the respective values of LB and NMAS remains the same. Intuitively, LB as a function of the area must be defined recursively, and such function respects the number of individuals of the most abundant species. From this intuition, it turns out, for BCI data, around 32 ha area LB increases by 1. As three species can be considered as the most abundant species and around 32 ha area the number of bins from 11 increases to 12, and also it implies that about 64 ha area, we must have 13 bins.

To test our results, we find the patterns from 2 (ha) to half of the area, then extrapolate to the total area to see how different the results are. Figure 7 shows how $C'_{10}(x)$ changes from 2 (ha) to 25 (ha), and then from that Figure 8 shows how the extrapolation of $C_{10}(x)$ changes from 2 (ha) to 50 (ha).

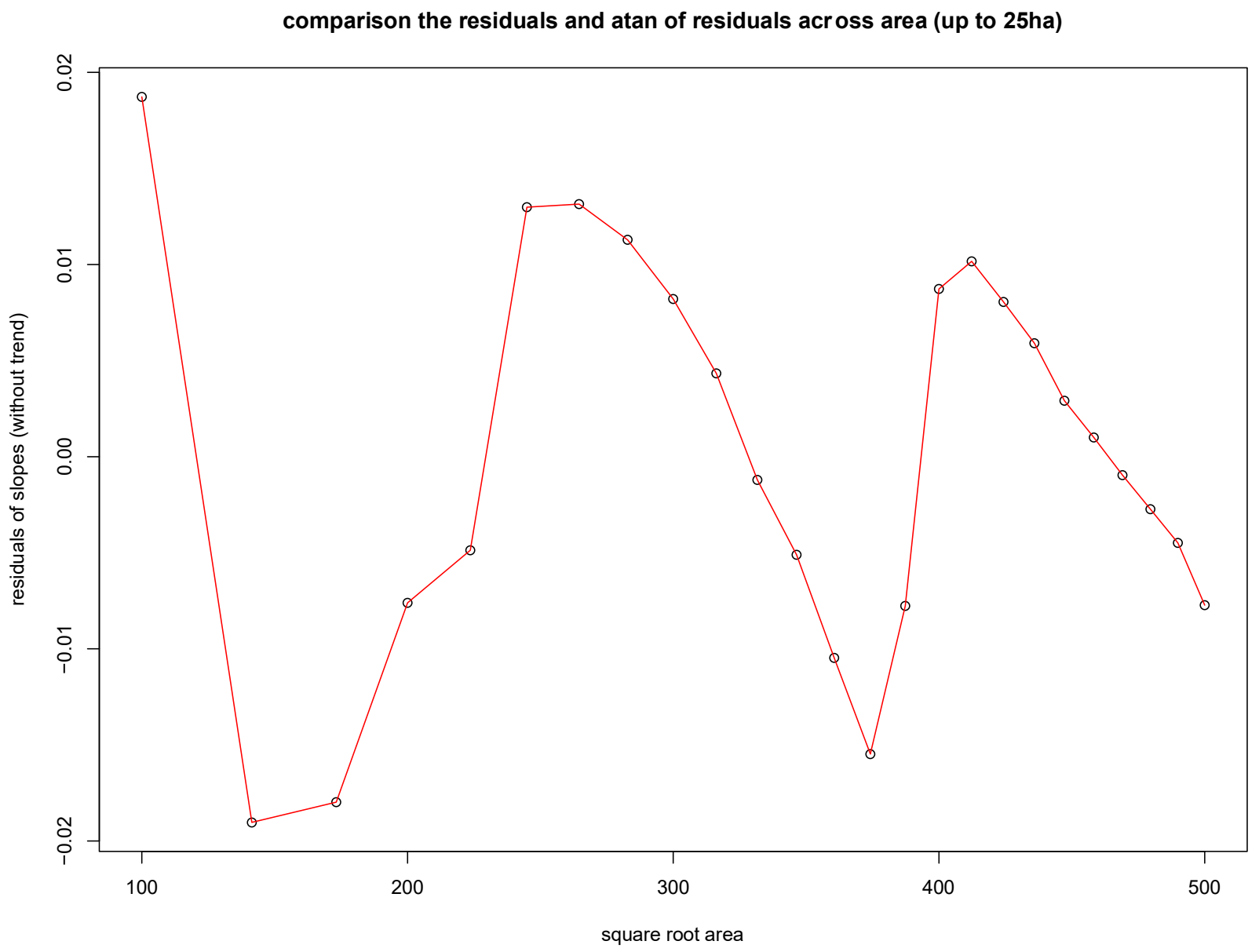

**Figure 7.** In the data from BCI, residuals of the linear regression of slopes of the log of 10th raw moment by the 10th distribution moment at the square root of area size from 2 (ha) to 25 (ha), circles are real data, and the red line is by considering residuals as arctan of itself.

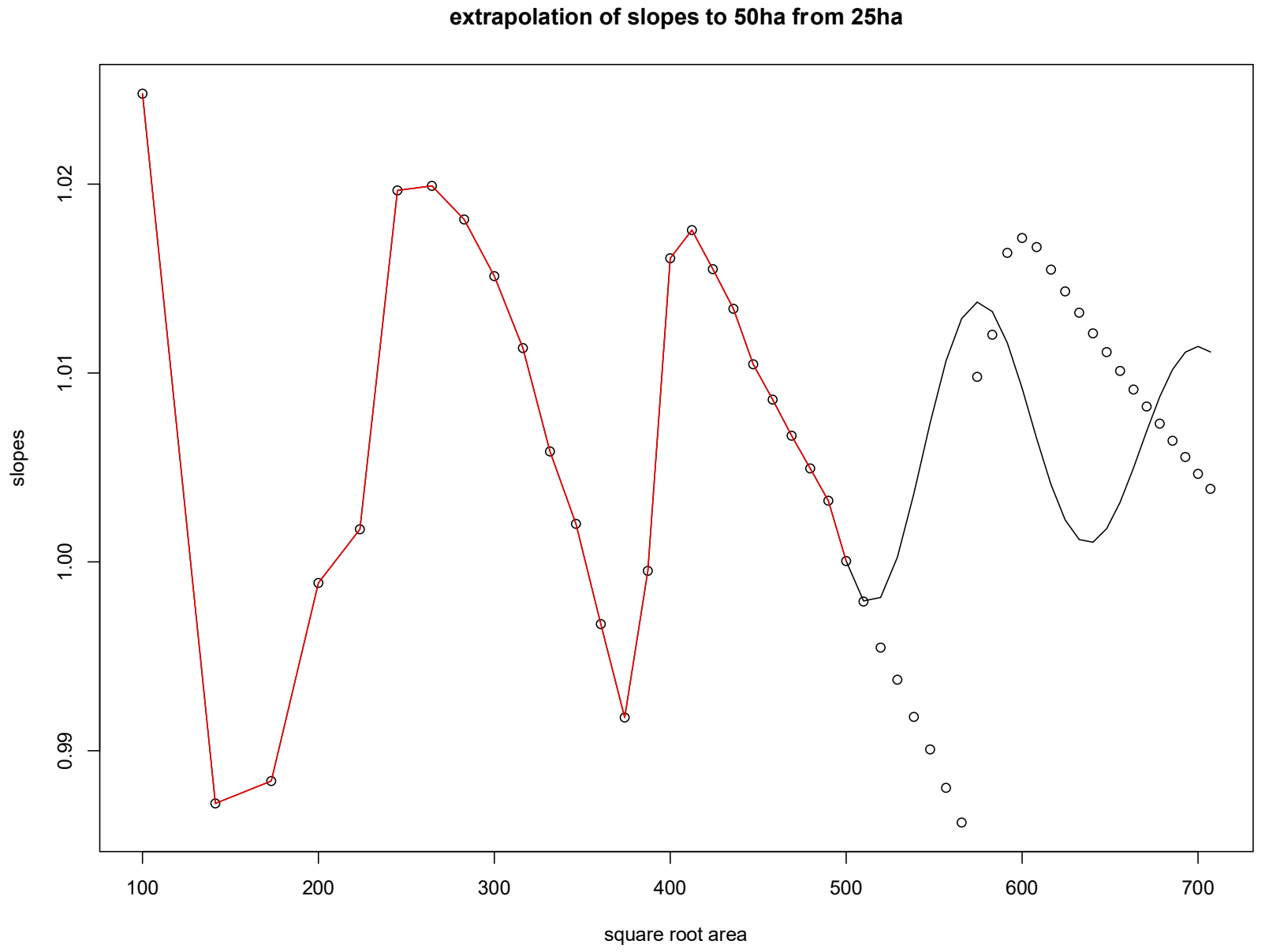

**Figure 8.** In the data from BCI, the slopes of the linear regression of the log of 10th raw moment by the 10th distribution moment at the square root of area size from 2 (ha) to 50 (ha), circles are real data, the red line is by considering residuals as arctan of itself, and the black line is the extrapolation by considering only 25 (ha) of the total area.

# 4. Reduce Computational Complexity

To find species abundance distribution from the given raw moments, we use Tchebychev moments and polynomials where their computations involve a large computational complexity. Here we use the recurrence relationship between Tchebychev polynomials to reduce their computational complexity. We introduce the matrix multiplier for computing Tchebychev moments from a given set of moments of different orders.

## 4.1. Recurrence Relation

Tchebychev polynomials are polynomials of the form

$$t_n(x) = \sum_{i=0}^{n} a_{i,n} x^i, \qquad (1)$$

where $a_{i,n}$'s are coefficients correspond to the maximum degree of the polynomial, and there is a formula to obtain the distribution values for them. After we obtained the values of moments of different orders, $M_i$'s, by using the following formula, Tchebychev moments of different orders can be obtained

$$T_n = \sum_{j=0}^{n} a_{i,n} M_i,$$

where $a_{i,n}$'s are the same as before. Then the species abundance distribution can be obtained as follows

$$f(x) = \sum_{i=0}^{n} T_i t_i(x).$$

The original formula of Tchebychev moments is used in Borda-de-Água et al. (2012); Borda-de-Água et al. (2017). When the moment order becomes large, the Tchebychev moments tend to exhibit numerical instabilities.

From Zhu et al. (2007); Chihara (1989); Wang and Wang (2006); Mukundan (2004), by using recurrence relation and since the polynomials are orthogonal, the computational complexity of finding the Tchebychev polynomials will be minimized. The Tchebychev polynomials up to degree $N$ follow the following recursive formula

$$t_n(x) = (\alpha_1 x + \alpha_2) t_{n-1}(x) - \alpha_3 t_{n-2}(x), \; n = 2, \cdots, N \text{ and } x = 0, 1, \cdots, N', \quad (2)$$

where the initial values are

$$t_0(x) = \frac{1}{\sqrt{N}} \text{ and } t_1(x) = \left(\frac{1-N+2x}{\sqrt{N(N^2-1)}}\right) \times \sqrt{3},$$

and

$$\alpha_1 = \left(\frac{2}{n}\right)\sqrt{\frac{4n^2-1}{N^2-n^2}}, \; \alpha_2 = \left(\frac{1-N}{n}\right)\sqrt{\frac{4n^2-1}{N^2-n^2}}, \text{ and } \alpha_3 = \left(\frac{n-1}{n}\right)\sqrt{\frac{2n+1}{2n-3}}\sqrt{\frac{N^2-(n-1)^2}{N^2-n^2}}.$$

By comparing equations (1) and (2) and by using symbolic calculations[1], the coefficients $a_{i,n}$'s for all values of $i$ and $n$ can be obtain and we have the following matrix representation of coefficients

$$\begin{bmatrix} a_{0,1} & 0 & 0 & 0 & \cdots & 0 \\ a_{0,2} & a_{1,2} & 0 & 0 & \cdots & 0 \\ a_{0,3} & a_{1,3} & a_{2,3} & 0 & \cdots & 0 \\ a_{0,4} & a_{1,4} & a_{2,4} & a_{3,4} & \cdots & 0 \\ \vdots & \vdots & \vdots & & \ddots & \vdots \\ a_{0,N} & a_{1,N} & a_{2,N} & a_{3,N} & \cdots & a_{N-1,N} \end{bmatrix},$$

[1] In a symbolic calculation of a function, we consider variable(s) of the function as a symbol (does not have numerical values) then we apply existing operations such as multiplication, summation, subtraction, power, and so on, in order to find the coefficients and powers of the variable(s).

whereby matrix multiplication and having the values of raw moments up to degree $N' - 1$, we have the values of Tchebychev moments up that order as follows

$$\begin{bmatrix} T_0 \\ T_1 \\ T_2 \\ T_3 \\ \vdots \\ T_{N'-1} \end{bmatrix} = \begin{bmatrix} a_{0,1} & 0 & 0 & 0 & \cdots & 0 \\ a_{0,2} & a_{1,2} & 0 & 0 & \cdots & 0 \\ a_{0,3} & a_{1,3} & a_{2,3} & 0 & \cdots & 0 \\ a_{0,4} & a_{1,4} & a_{2,4} & a_{3,4} & \cdots & 0 \\ \vdots & \vdots & \vdots & & \ddots & \vdots \\ a_{0,N'} & a_{1,N'} & a_{2,N'} & a_{3,N'} & \cdots & a_{N'-1,N'} \end{bmatrix} \times \begin{bmatrix} M_0 \\ M_1 \\ M_2 \\ M_3 \\ \vdots \\ M_{N'-1} \end{bmatrix}.$$

Note that the values of $x$ are $0, 1, \dots,$ the maximum number of bins. These calculations will reduce the computational complexities for evaluating both Tchebychev polynomials and Tchebychev moments.

## 5. Additional Error Reduction

The maximum number of bins is 1000, and Figure 9 shows the orthonormal Tchebychev polynomials up to degree 5.

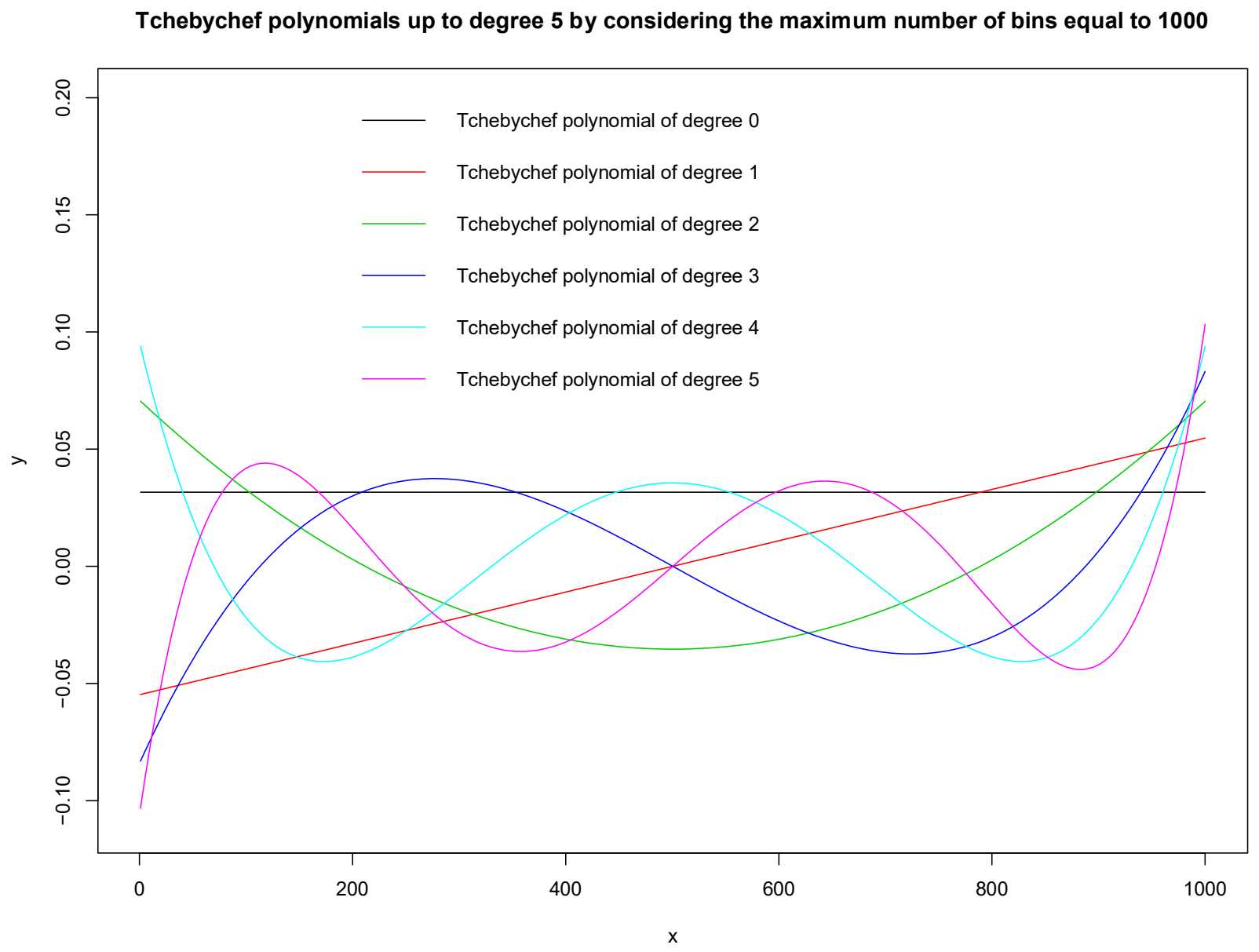

**Figure 9.** We consider the maximum number of bins is 1000, and then we draw the Tchebychev polynomials up to degree 5.

In Figure 9, if the maximum number of bins is very law, generally, we have a very low number of bins in species abundance distributions. Instead of smooth shapes for higher degree polynomials, we start to have the stagger shapes, which is another problem for using Tchebychev polynomials and moments. To handle this last issue, note that for a given data, the species abundance distribution is a distribution function that is defined over the interval $[0, M]$ where $M$ is the maximum number of bins obtained from data, and so we may assume that the values of the distribution function outside of this interval is equal to zero. Under this assumption, the maximum number of bins can be considered much larger, but raw moments' values will not change. In other words, we consider the real species abundance distribution as a part of the bigger one, which involves more bins with the number of species 0. Note that if we use a very large number for the number of bins, we must use more moments in our calculations to get a better result. By using the small number of moments, intuitively, we will have a shade at the beginning and then a straight line over the x-axis. To handle this problem, we suggest that we must add higher-order moments by increasing the number of bins. Hence, we are almost open-handed to use a larger value for the number of bins, and also, the order of moments can be higher. In the original case, the number of moments involved in the calculations, in theory, is at most the maximum number of bins minus 1, whereby our observation for using moments of high order the staggering behavior starts to appear. But now, in the current version, after we find the Tchebychev polynomials and moments then by the formula

$$f(x) = \sum_{i=0}^{n} T_i t_i(x),$$

we obtain the distribution, but the values of $x$ take place in the interval of zero and the new number of bins that we newly considered. We only need to use the restriction of this function to the interval of interest, which is from zero up to the maximum number of bins.

### 5.1. Analyzing Errors

Here we consider BCI data, including all the individuals of all species in the 50 ha area. In Figure 10, we considered the number of bins as 20, 30, 40, …, 800. We check how much we need to consider as the maximum order of raw moments in our calculations.

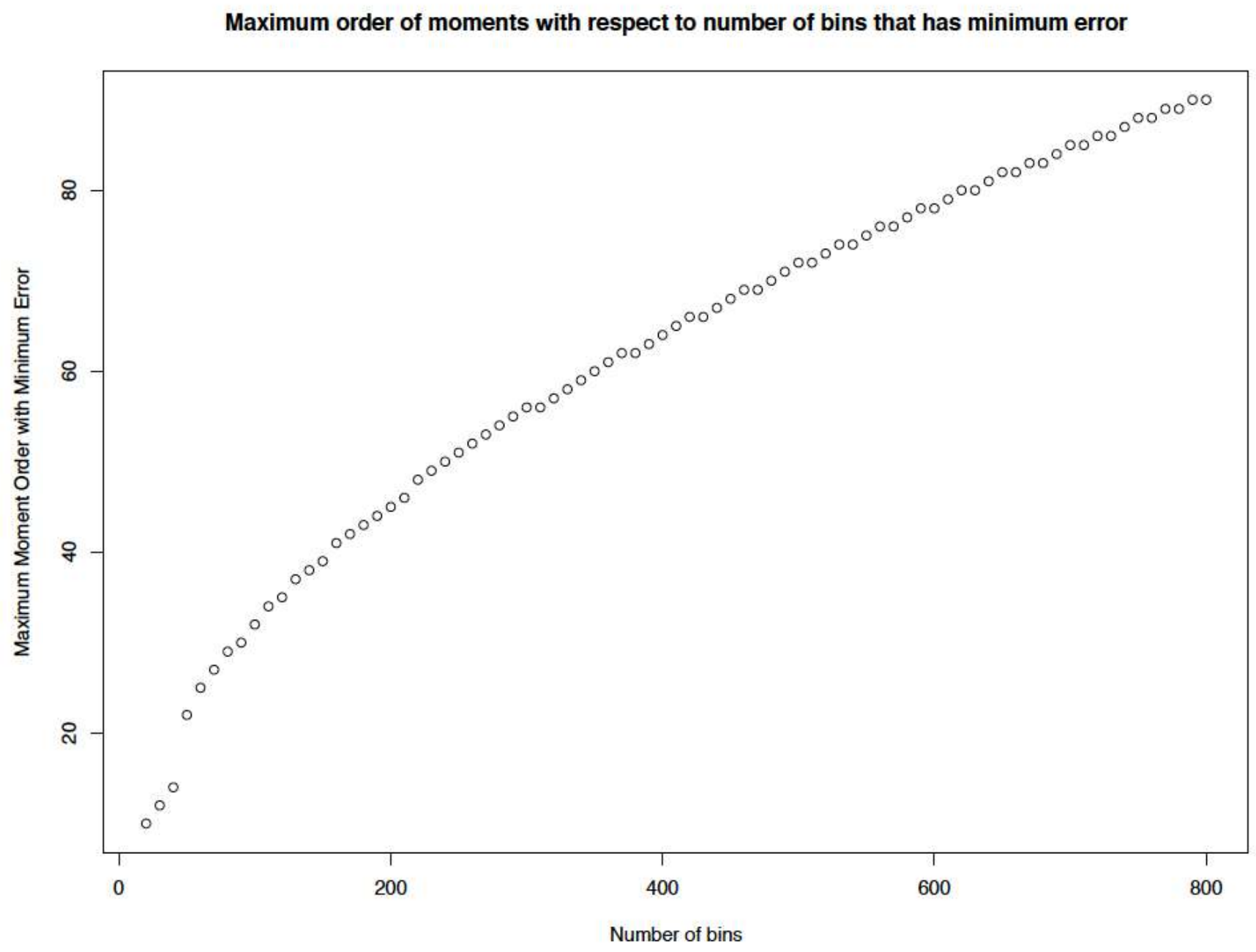

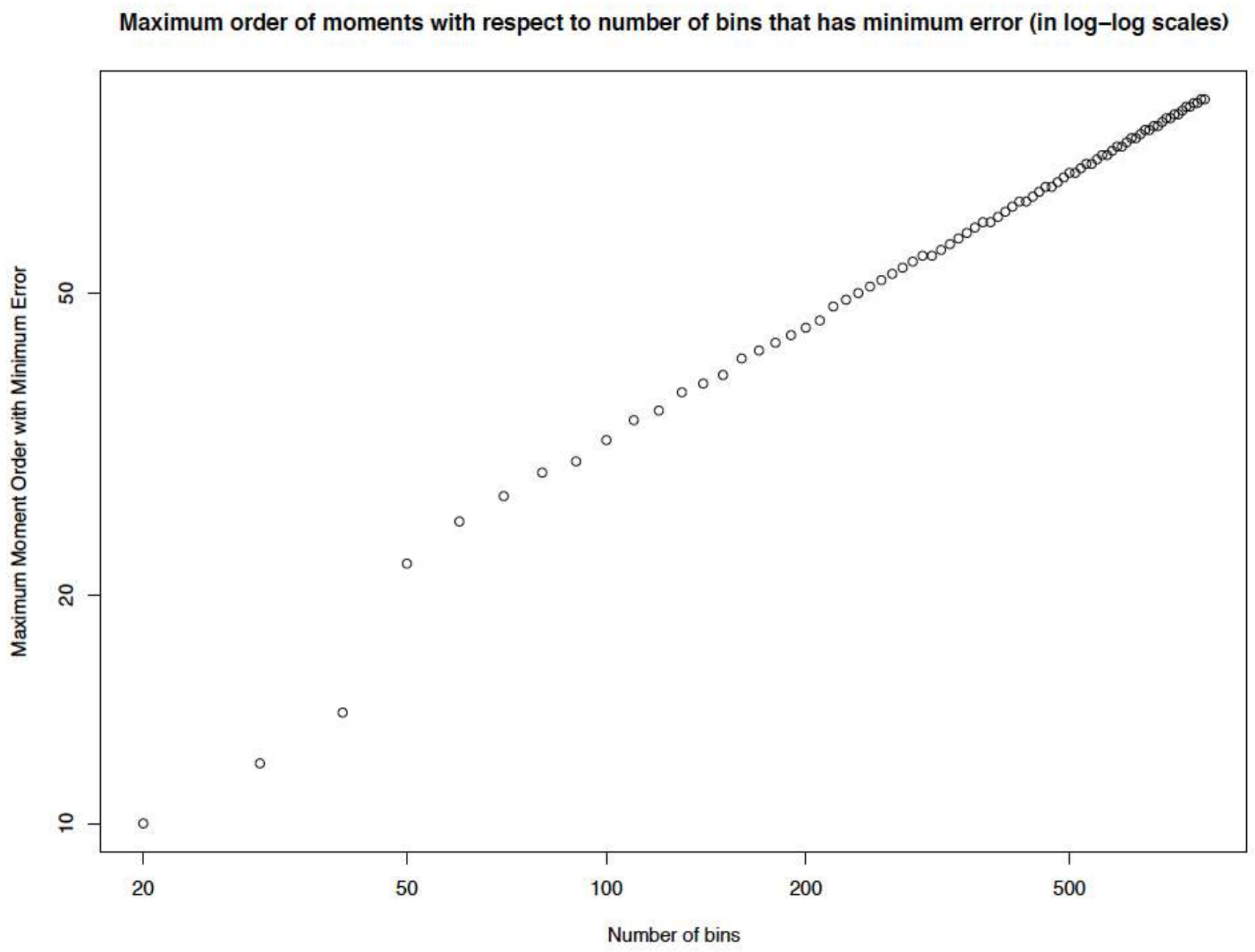

**Figure 10.** We consider additional bins up to 800 for data from BCI and find the proper maximum order for moments such that the sum of the square of differences between real SAD and the one that can be obtained from moments is minimum. We plot the moment order corresponds to the number of bins. The first one is in arithmetic, and the second one is in the log-log scale.

As we can see in the log-log scale, a linear pattern appears. Now to see how accurate we are, we plot the errors (the sum of the square of differences between real SAD and the one obtained by considering raw moments) of corresponding maximum moments order as a function of the number of bins. As we can see in Figure 11, the errors are decreasing.

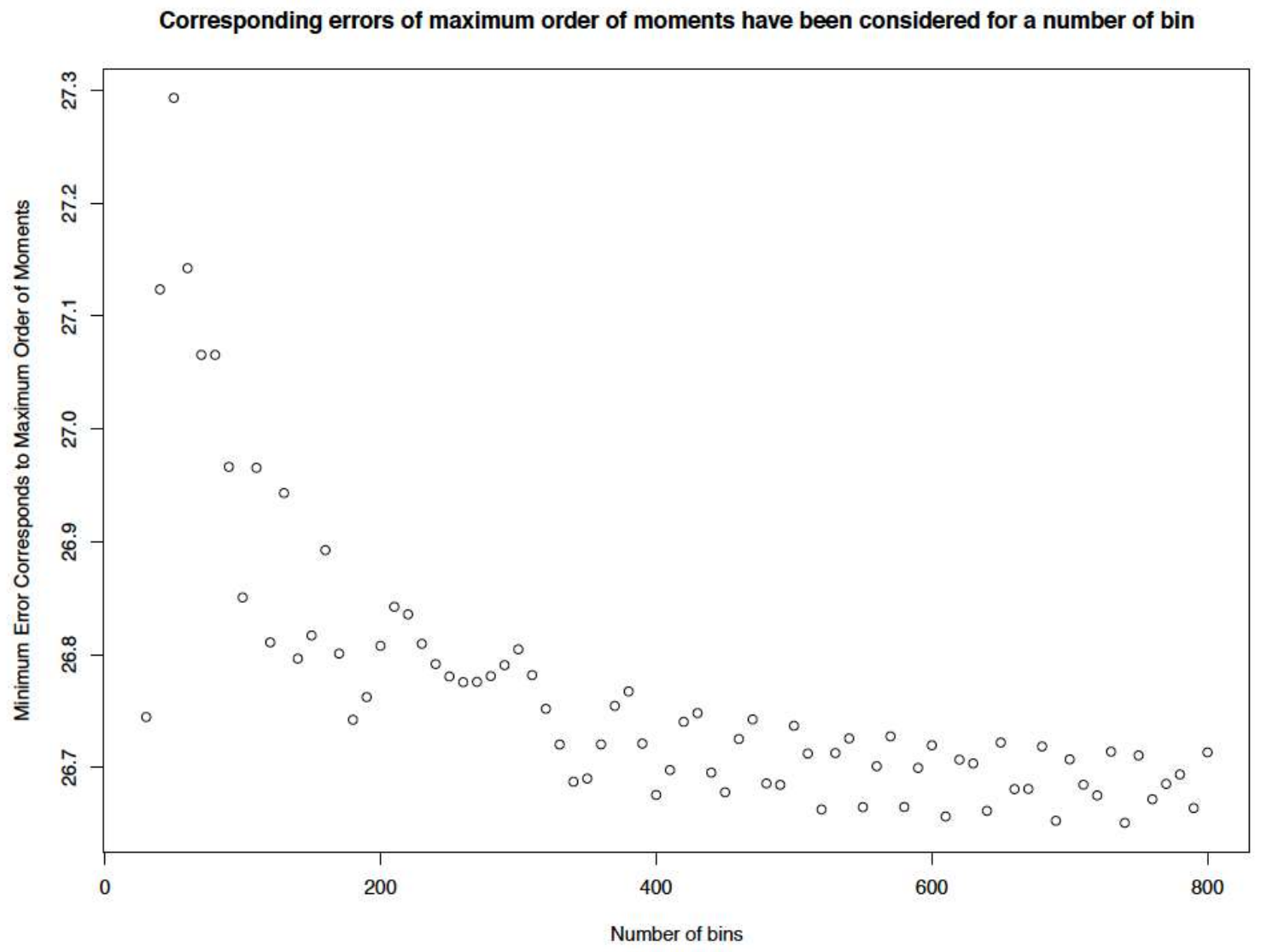

**Figure 11.** We consider additional bins up to 800 for data from BCI and find the proper maximum order for moments such that the sum of the square of differences between real SAD and the one that can be obtained from moments is minimum. We plot the minimum errors correspond to the number of bins.

Note that we are only able to find moments up to degree 250. Figure 12 shows how the moments' orders are with respect to the number of bins that the minimum errors occur by considering the number of bins equal to 500, 510, 520, …, 4000. Then we plot it in log-log scales, and finally, we show how the corresponding minimum errors with respect to the moments of suitable order change as a function of the number of bins.

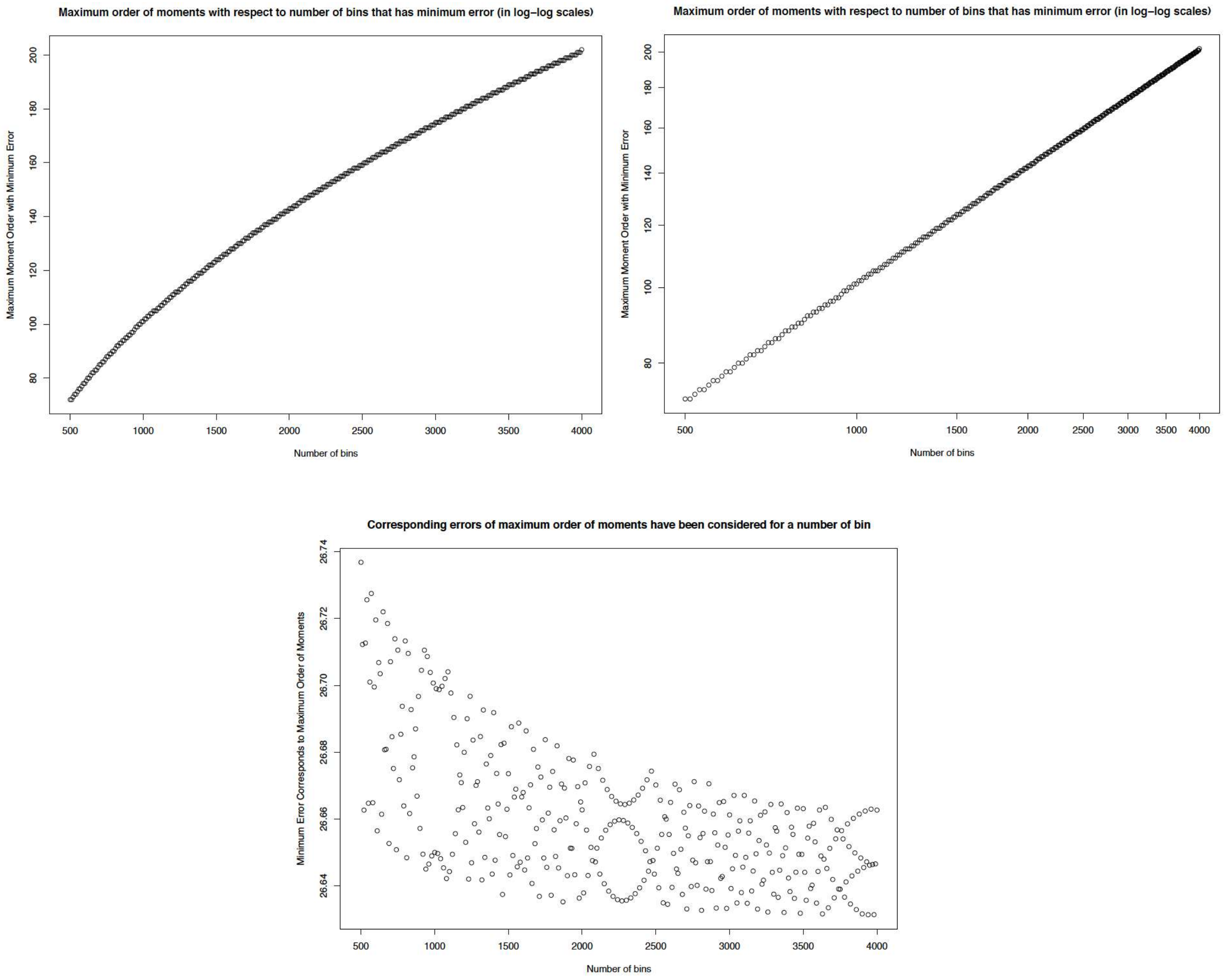

**Figure 12.** We consider additional bins up to 4000 for data from BCI and find the proper maximum order for moments such that the sum of the square of differences between real SAD and the one that can be obtained from moments is minimum. We plot the moment order corresponds to the number of bins. The first one is in arithmetic, and the second one is in a log-log scale. We plot the minimum errors correspond to the number of bins.

Recall that moments have information about the species abundance distribution, so it is necessary to involve moments of higher orders in the calculations. Our method is involved in more details, which are obtained from moments in the result. The linear pattern of maximum moment orders by the number of bins suggests that, as we can only obtain the moments up to order 250, the number of bins we can use in our arguments should not reach 6168 bins. With respect to BCI data, by letting $N$ to be number of bins, the maximum order of moment we need to consider follows the following formula (with $R^2 = 0.9999325$):

$$\exp(0.4977 \log N + 1.1779).$$

## 6. Theory in Example

Our example is the data of tropical forest BCI, which consists of the number of individuals with 10cm dbh per species. In Figure 13, we compare the results of evaluating species abundance distribution from raw and distribution moments. For distribution moments, we use 12 moments, and for raw moments we use 12 and 11 moments. By error, we mean the sum of absolute differences.

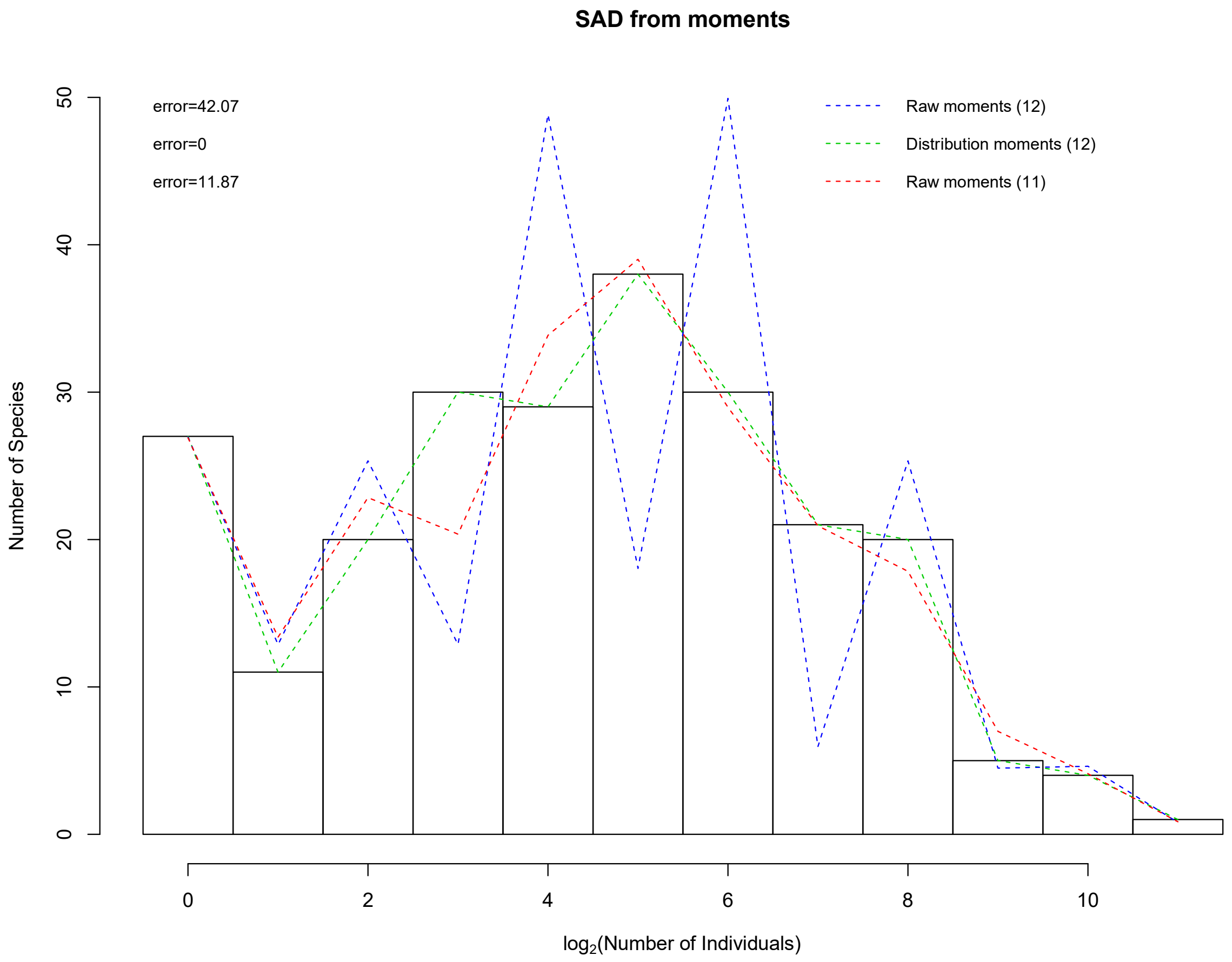

**Figure 13.** SAD for BCI, all individuals with at least 10cm dbh are considered. We plot over it the SADs obtained from moments; Green and dashed line is by considering distribution moments up to degree 12; Dark Blue and dashed line is by considering raw moments up to degree 12; Red and dashed line is by considering raw moments up to degree 11.

To better visualize the differences in Figure 14, we drop the case with the use of 12 raw moments.

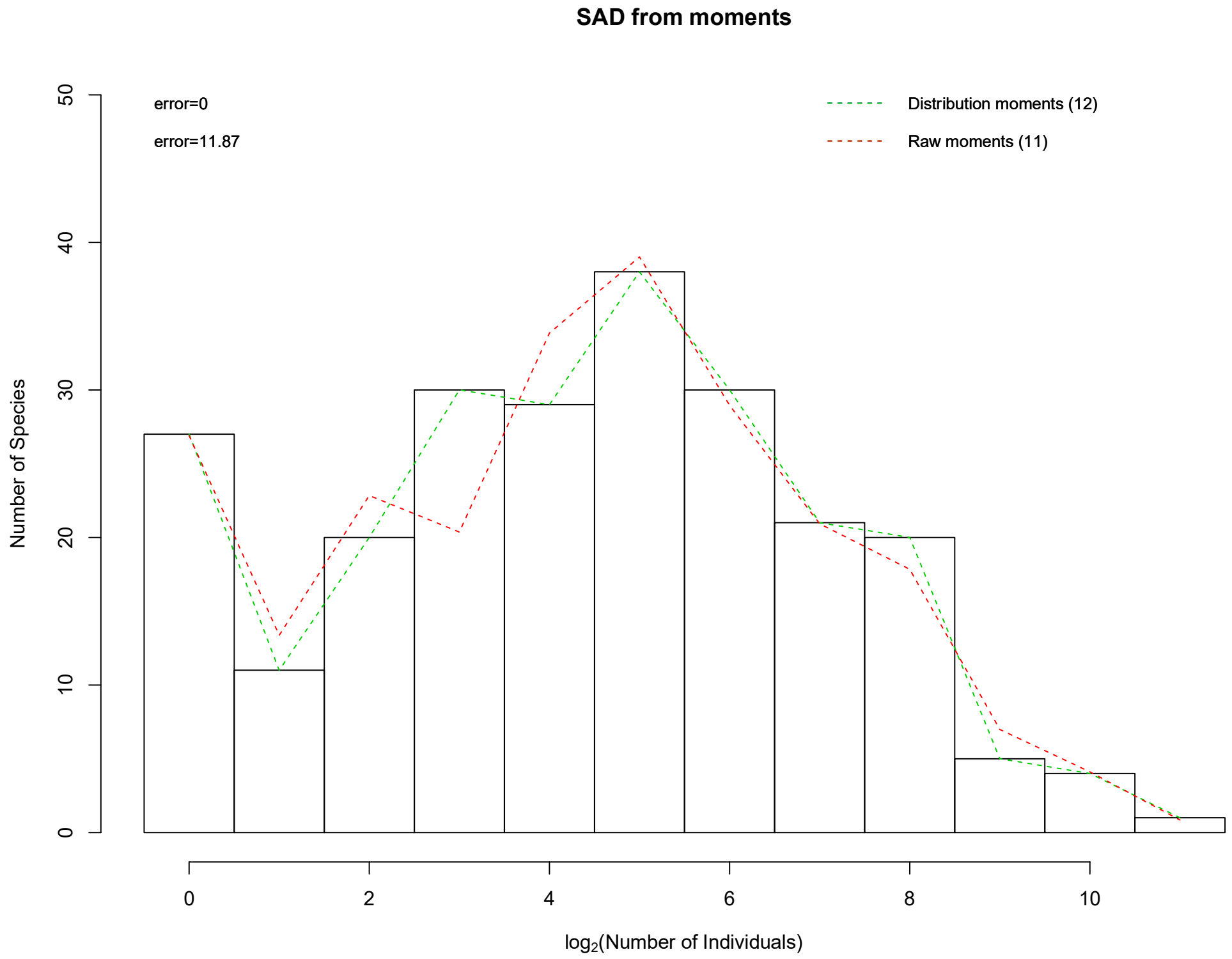

**Figure 14.** SAD for BCI, all individuals with at least 10cm dbh are considered. We plot over it the SADs obtained from moments; Green and dashed line is by considering distribution moments up to degree 12; Red and dashed line is by considering raw moments up to degree 11;

In Figure 15 for the sample data BCI, we extrapolate the species abundance distribution for the area of 50 ha by only assuming a sub-area of size 25 ha by using extrapolation of moments up to order 10. For distribution moments, we use 7 moments, and for raw moments we use 7 and 6 moments respectively for non-linear and linear; these values correspond to the values that have minimum errors.

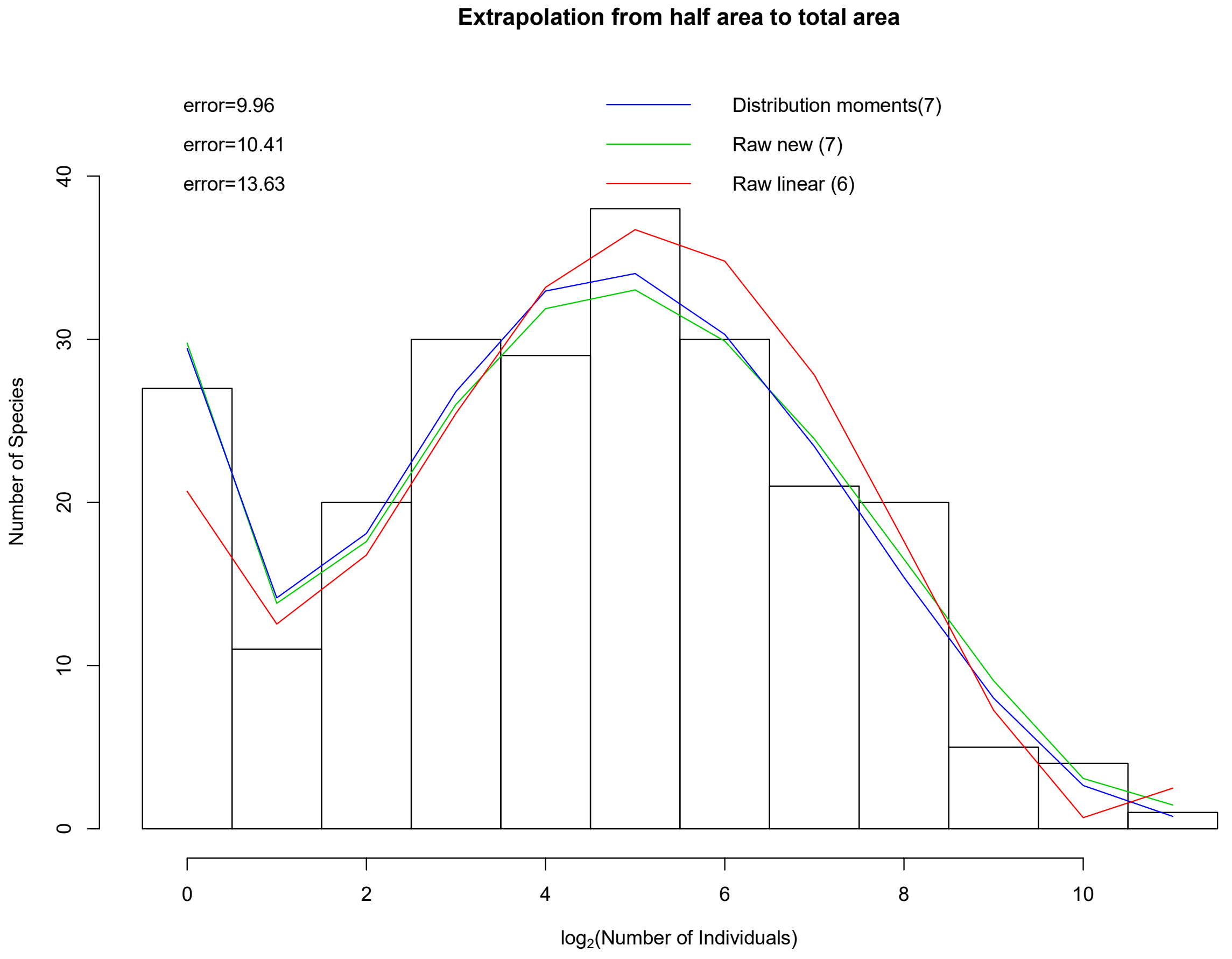

**Figure 15.** Extrapolating SAD for BCI, all individuals with at least 10cm dbh are considered by extrapolating the moments. And by only considering data from 25 (ha) area. The histogram is the real data. We plot over it the SADs obtained from extrapolation of moments to 50 (ha). The red line is by extrapolating raw moments under consideration linear relation, and moments up to degree 6 is used. The green line is by extrapolating raw moments under consideration of non-linear function, and moments up to degree 7 are used. The blue line is by extrapolating distribution moments, and moments up to degree 7 are used.

In Figure 15, we showed how the results are different. By considering the linear version of extrapolation and the non-linear interpretation for BCI with individuals with 10cm (dbh) and extrapolation to 50 ha from a subarea of 25 ha.

In Figure 16, for the sample data BCI, we extrapolate the species abundance distribution to 50 ha by using the sub-area of size 25 ha and using extrapolation of distribution moments up to degree 10. We considered the case of adding an extra bin and then consider 8 instead of 7 moments. Note that by involving more moments, we add additional information about the SAD.

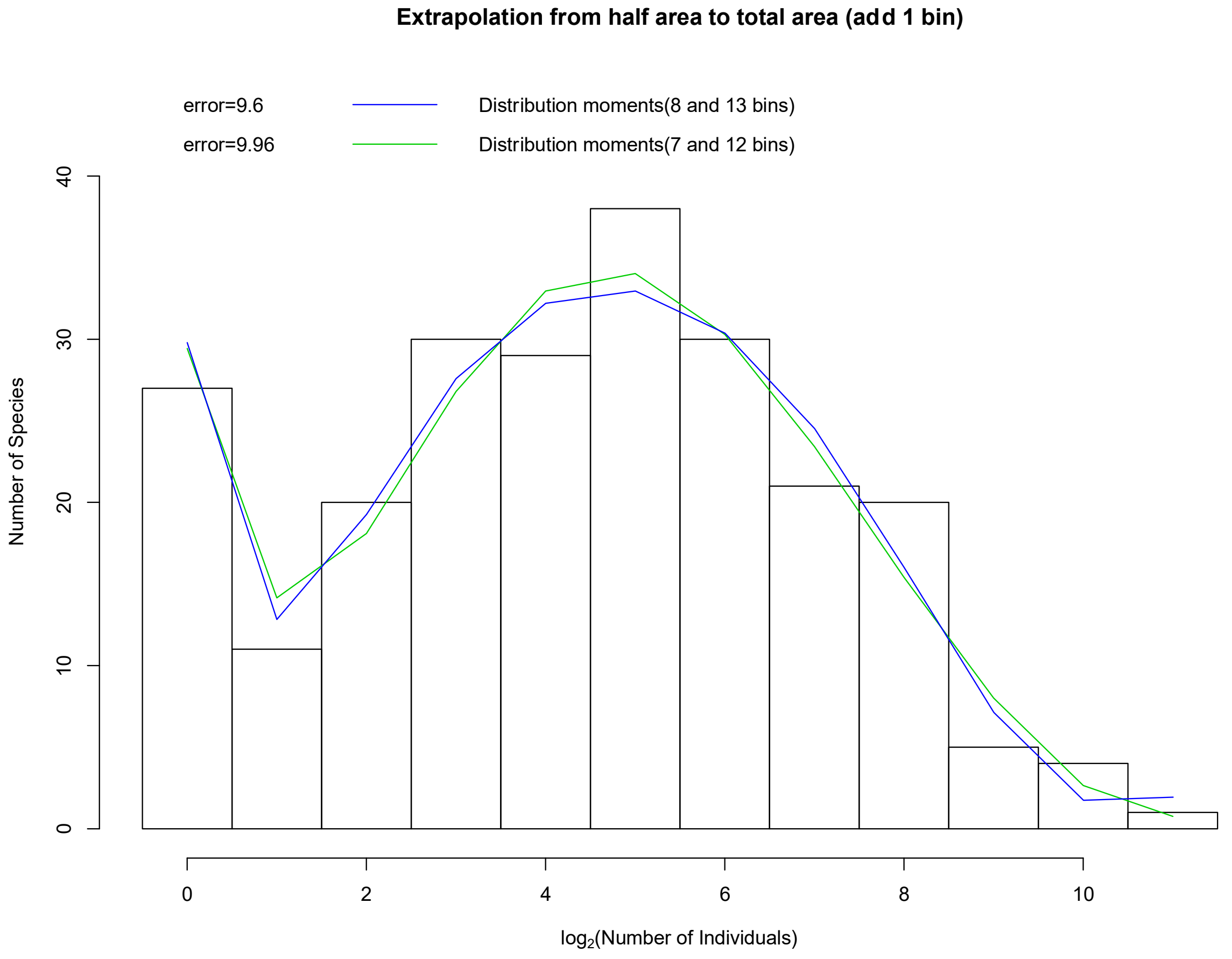

**Figure 16.** Extrapolating SAD for BCI, all individuals with at least 10cm dbh are considered by extrapolating the moments. And by only considering data from 25 (ha) area. The histogram is the real data. We plot over it the SADs obtained from extrapolation of distribution moments to 50 (ha). The green line is by extrapolating moments under consideration that the number of bins is 12 and moments up to degree 7 are used. The blue line is by extrapolating distribution moments considering that the number of bins is 13 and moments up to degree 8 are used.

# Conclusion

The accurate fit for species abundance distribution can be obtained using the distribution moments in a fixed area. We start by considering distribution moments and formulate it as a function of raw moments. We provide a way of extrapolation of distribution moments for larger scales through extrapolation of raw moments.

## Acknowledgments

This research was partially supported by Fundação para a Ciência e Tecnologia (PTDC/BIA-BIC/5558/2014) as the BPD project “species diversity as a function of spatial scales” with reference number ICETA-2016-43. The BCI forest dynamics research project was founded by S.P.

Hubbell and R.B. Foster and is now managed by R. Condit, S. Lao, and R. Perez under the Center for Tropical Forest Science and the Smithsonian Tropical Research in Panama. Numerous organizations have provided funding, principally the U.S. National Science Foundation, and hundreds of field workers have contributed. We also acknowledge R. Foster as plot founder and the first botanist able to identify so many trees in a diverse forest; R. Pérez and S. Aguilar for species identification; S. Lao for data management; S. Dolins for database design; plus hundreds of field workers for the census work, now over 2 million tree measurements; the National Science Foundation, Smithsonian Tropical Research Institute, and MacArthur Foundation for the bulk of the financial support.